\documentstyle [12pt, a4wide, psfig] {article}
\begin{document}
\begin{titlepage}
\begin{center}
\vspace{2cm}
\LARGE
Detection of Strong Evolution in the Population of Early-Type Galaxies\\
\vspace{1cm}
\large
Guinevere Kauffmann$^{1}$, St\'{e}phane Charlot$^{2}$ \& Simon D.M. White$^{1}$
\\
\vspace{0.5cm}
\small
{\em $^1$Max-Planck Institut f\"{u}r Astrophysik, D-85740 Garching, Germany} \\
{\em $^2$ Institut d'Astrophysique du CNRS, 98 bis Boulevard Arago, F-75014
Paris, France} \\
\vspace{0.8cm}
\end{center}
\normalsize
\begin {abstract}

The standard picture holds that giant elliptical galaxies
formed in a single burst at high redshift.
Aging of their stellar populations subsequently caused them to fade
and become redder. The Canada-France Redshift Survey provides a sample
of about 125 galaxies with the luminosities and colours of passively
evolving giant ellipticals and with $0.1 < z < 1$. This sample is
inconsistent with the standard evolutionary picture with better than 99.9\%
confidence.
The standard Schmidt test gives $\langle V/V_{max}\rangle = 0.398$ when
restricted to objects with no detected star formation, and
$\langle V/V_{max}\rangle = 0.410$ when objects with emission lines
are also included.  A smaller sample of early-type galaxies selected from the
Hawaii Deep Survey gives equally significant results.
With increasing redshift a larger and larger
fraction of the nearby elliptical and S0 population must drop out of
the sample, either because the galaxies are no longer single units or
because star formation alters their colours. If the remaining fraction
is modelled as $F=(1+z)^{-\gamma}$, the data imply $\gamma=1.5\pm 0.4$.
At $z=1$ only about one third of nearby bright E and S0 galaxies were
already assembled and had the colours of old passively evolving
systems. We discuss the sensitivity of these results to the
incompleteness corrections and stellar population models we have adopted.
We conclude that neither is uncertain enough to reconcile the
observations with the standard picture. Hierarchical galaxy formation
models suggest that both merging and recent star formation play a role
in the strong evolution we have detected.

\end {abstract}
\vspace {0.8 cm}
\end {titlepage}

\section {Introduction}
In the standard model of elliptical galaxy formation first proposed by Tinsley
\&
Gunn (1976), all stars in the galaxy form in an initial burst at high redshift
and the galaxy's luminosity subsequently evolves {\em passively} as the more
massive stars evolve off the main sequence.
This leads to luminosity evolution with cosmic time that can be parametrized as
\begin {equation} L(t)= L(t_0) [ (t-t_f)/(t_0-t_f)] ^{-1 + \theta x}, \end
{equation}
where $t_f$ is the time of formation,
$x$ is the slope of the initial mass function ($x=1.35$ for a Salpeter
mass function), and $\theta \simeq 0.26$ is the slope of the mass-main sequence
lifetime
relation (see Phillipps 1993). Spectral synthesis techniques show that simple
passive
evolution models provide good fits to the colours and spectral properties of a
class
of galaxies conventionally called {\em early-type}, which include the
ellipticals and the S0s.

With the superior imaging capability of the Hubble Space Telescope, it has
recently
become possible to identify early-type galaxies at high redshift purely
on the the basis of their morphologies (see for example Driver et al 1995,
Abraham et al 1996). There is an emerging consensus that the colours and
luminosities
of early-type galaxies in  rich clusters are
consistent with simple passive evolution models and with a standard value of
the
IMF slope (Aragon-Salamanca et al 1993, Dickinson 1995, Bender, Ziegler \&
Bruzual 1996,
Van Dokkum \& Franx 1996, Schade et al 1996, Pahre, Djorgovski \& de Carvalho
1996, Ellis et al 1996).
It appears that the luminosity evolution long
predicted as a consequence of  stellar aging has
finally been detected. These studies do not prove, however, that the early-type
galaxy
population {\em as a whole} formed at high redshift and evolved passively.
A test of this hypothesis requires a large redshift survey of early-type
galaxies
that is complete to faint limiting magnitudes.

Morphological classification of early-type galaxies has been carried out to
limiting
magnitudes of $I=21$ and $I=25$ in the Medium Deep and the Hubble Deep Field
surveys
respectively, but redshifts for a complete sample of these galaxies are lacking
at present.
The Canada France Redshift Survey (CFRS) is one of the deepest large redshift
surveys
carried out to date
(Lilly et al 1995a, Le F\`{e}vre et al 1995,
Hammer et al 1995, Crampton et al 1995, Lilly et al 1995b).
As we will show, it provides a sample of about 125 I-band selected galaxies
with the luminosities and colours of passively evolving giant ellipticals and
with redshifts lying in the range 0.1 to 1. Morphological classifications
for these objects are not yet available. The Hawaii Deep Fields Spectroscopic
Survey
(Cowie et al 1996) provides us
with an independent K-band selected sample of 28 early-type galaxies.
The great advantage of working with
magnitude-limited redshift surveys is that a given evolutionary hypothesis
may be tested directly in a way which is independent of the galaxy luminosity
function.
This was first illustrated in a seminal
paper by Schmidt (1968), who introduced the so-called $V/V_{max}$ test and used
a flux-limited sample
of only 33 QSOs to show that the population has evolved strongly with redshift.
The test is particularly powerful in the present context because it avoids the
need to assume
anything about the highly uncertain {\em local} luminosity function of
early-type
galaxies (compare the very different results of Loveday et al (1992) and Marzke
et al (1994)).

In this Letter, we apply the Schmidt $V/V_{max}$ test to early-type galaxies
selected
by colour from the CFRS and Hawaii surveys.
We show that these samples are both inconsistent
with the simple passive evolution picture with better than 99.7\% confidence.

\section {Selecting Early-type Galaxies in the CFRS}

Ellipticals and S0s are the reddest objects in the local Universe.
To define a colour threshold for
separating early from late-type galaxies, we make use of new
population synthesis models by Bruzual \& Charlot (1996),
which include the effects of metallicity (see also Charlot 1996).
For our standard model, we assume that all early-type galaxies form in a single
burst of duration
0.1~Gyr at $z=5$ and have a
Salpeter IMF with upper and lower cutoffs at 100$M_\odot$ and 0.1$M_\odot$.

Ideally, our adopted colour boundary should be blue enough to include all
elliptical and S0
galaxies, yet red enough to exclude spiral galaxies of type Sa or later. In
practice
galaxies of given type show a substantial spread in colour, so no perfect
boundary exists.
We adopt a colour threshold corresponding to a passively evolving model with
50\% solar
metallicity. We have chosen this model because it gives the best separation
between early
and late-type galaxies in local photometric surveys. We find that under this
criterion,
88\% of the galaxies in the Bower, Lucey \& Ellis (1992)
sample of 66 ellipticals and S0s in  the Coma and Virgo clusters would be
included in our
sample, while 84\% of the galaxies in the Visvanathan (1992) survey of 177
luminous Sa-Sc spirals towards the Great Attractor region would be rejected.

Figure~1 shows the resulting $(V-I)_{AB}$  redshift relation superposed on the
CFRS data.
We identify 98 early-type galaxies down to a limiting magnitude
of $I=22.5$, $17\%$ of the total sample of 591 galaxies with secure redshifts.
In the Hawaii Survey, galaxies are selected according to $B-K$ colour. We
identify
23 early-type galaxies to $K=19.5$, again about one fifth of the total sample.
Note that these fractions agree rather well with the $20-25 \%$ of galaxies
brighter than
I=22.5 in the HST Medium Deep Survey which have early-type morphology according
to
Driver et al (1995) and Abraham et al (1996).

\begin{figure}
\centerline{
\psfig{figure=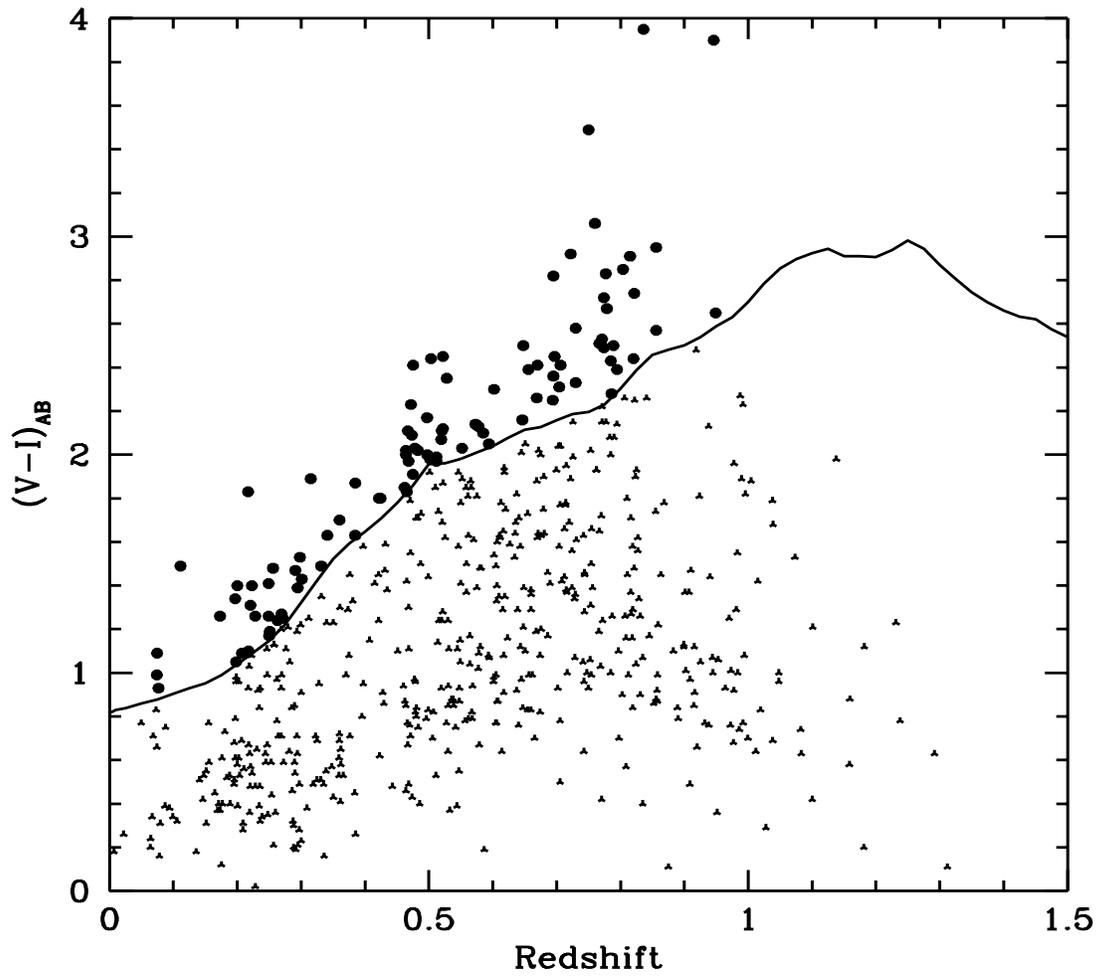,width=15cm,height=15cm}
}
\caption{\label{fig1}
 The division of the CFRS galaxies into early and late-types according
to the colour-redshift criterion discussed in the text.
Solid circles represent  the ellipticals and S0s and stars represent the
spirals and irregulars.}
\end {figure}

In addition to information about colours, magnitudes and redshifts, the CFRS
catalogue
contains a list of the main features seen in the spectrum of each galaxy.
Evidence for
ongoing star formation is provided by the detection of O[II] emission at
3727 \AA. Such emission is almost never seen in nearby E/S0 galaxies.
If we exclude galaxies with detected O[II], our sample
is reduced to 72 galaxies. In figure 2, we plot redshift and
absolute magnitude histograms for these samples. Note that the
absolute magnitudes are present-day values,
assuming that each galaxy evolves passively
from the time of detection until the present day.
Galaxies with detectable
O[II] emission tend to lie at at higher redshift and to be intrinsically
fainter than
those with no such emission. However, almost all objects in the sample have
absolute magnitudes
typical of giant ellipticals.

It is important to realise that the CFRS survey is only just over 80\%
complete. The
Hawaii Survey is 87\% complete to K=19.5. Omission of
the galaxies without secure redshifts would introduce significant bias into our
analysis,
since  many of them
are red and lie close to the magnitude limit of the survey.
It is thus important that we correct for incompleteness as well as possible and
investigate
the uncertainties which this correction introduces.

Our ``standard'' correction procedure is as follows and is similar to one
described by
Lilly et al (1995b).
We assume that each galaxy without a measured redshift has the same
distribution in $z$ as
galaxies of the same apparent magnitude and colour which do have secure
redshifts.
In practice, we define intervals in $I$ and  $V-I$ (or $K$ and $B-K$)
centred on the galaxy and
adjust them so that they contain 20 galaxies with redshifts.
We randomly select one of these galaxies
and assign its redshift to our candidate. If the galaxy then falls above the
colour threshold shown in figure 1, it is classified as early-type. As shown in
figure 2 for
the CFRS data,
this procedure adds about 28 objects to the sample, about 20\% of all the
galaxies without
secure redshifts.
Most lie close to $I=22.5$
and are redder than the majority of  galaxies of the same apparent magnitude
which do have secure redshifts. Their redshift and absolute magnitude
distributions
are thus biased towards higher values of $z$ and somewhat fainter luminosities.

\begin{figure}
\centerline{
\psfig{figure=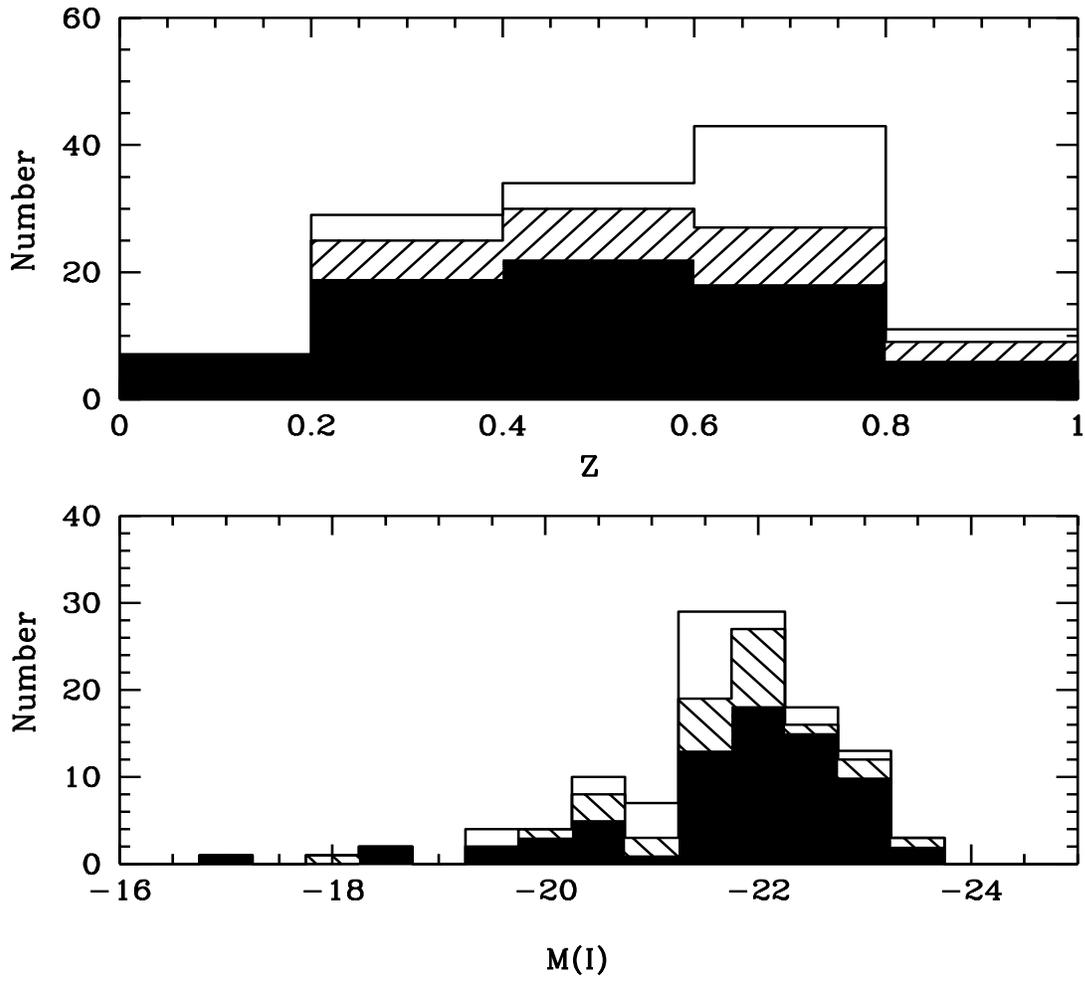,width=15cm,height=15cm}
}
\caption{\label{fig2}
Redshift and absolute magnitude histograms for early-type galaxies
selected from the CFRS catalogue. The solid area is for galaxies with no
detected
O[II] emission and the hatched area for galaxies with O[II] emission.
The unfilled area represents galaxies included by the incompleteness correction
described in section 2.}
\end {figure}

\section {The Schmidt Test}
The ratio $V/V_{max}$ is a measure of the position of a source within the total
volume $V_{max}$ where it could have been included in the sample.
To determine $V_{max}$, one must know the intrinsic luminosity of
the source, and hence the distance at which it would fall below the limiting
magnitude of
the survey, assuming that it evolves in luminosity according to the passive
evolution
model described in section 2. When applied to the CFRS data, one must also take
account of the fact that
no galaxies brighter than $I=17.5$ are included in the survey.
If passive evolution is the correct model, $V/V_{max}$
should be uniformly distributed between 0 and 1 and
$\langle V/V_{max} \rangle$ should scatter around 0.5, with a variance only
dependent on sample size.
In this Letter, we calculate all magnitudes and
volumes assuming that $q_0=0.5$ and $H_0= 50$ km s$^{-1}$ Mpc$^{-1}$. We have
checked that changing
$q_0$ has almost no effect on our results.

In figure 3, we show the $V/V_{max}$ distribution for early-type galaxies
selected
by colour from the CFRS catalogue. The solid histogram shows the result
obtained if
galaxies with O[II] emission are excluded from the sample, while the hatched
area
shows the contribution from galaxies with such O[II] emission.
Finally, the unfilled area shows the $V/V_{max}$ distribution of the galaxies
included
by the incompleteness correction of section 2.

The $V/V_{max}$ distribution is clearly skewed to values less than 0.5. We
obtain
$\langle V/V_{max} \rangle = 0.398$ when objects with
detected star formation are excluded and
$\langle V/V_{max} \rangle =0.410$ when they are included.
For a uniform distribution, the expected standard deviations in $\langle
V/V_{max} \rangle$ for samples
of these sizes are 0.029 and 0.026 respectively. The observational data are
thus inconsistent
with the standard model of passive evolution with better than 99.9\%
confidence.
It seems that with increasing redshift, a larger and
larger fraction of the nearby elliptical/S0 population must drop out of the
sample, either
because the galaxies are no longer single units or because star formation
alters
their colours. We can parametrize the evolution of the remaining fraction
as $F= (1+z)^{-\gamma}$ and determine the value of $\gamma$
required for the $V/V_{max}$  distribution to come out uniform. We find that
$\gamma =1.7 \pm 0.4$ for the sample without O[II] emission and that
$\gamma=1.5 \pm 0.4$
for the full sample (see the lower panel of figure 3). This density evolution
is
quite dramatic! It implies that at $z=1$ only about one third of nearby bright
E and S0 galaxies were already assembled and had the colours of old, passively
evolving stellar systems.

For the 23 early-type galaxies identified in the Hawaii data, we obtain
$\langle V/V_{max}
\rangle = 0.317$ The expected variance for this sample is 0.060.
Thus even though this sample is much smaller, the deviation of $\langle
V/V_{max} \rangle$
from 0.5 is almost as significant as in the CFRS data. The smaller value of
$\langle V/V_{max} \rangle$
reflects the fact that passively evolving galaxies could in principle be
detected to higher redshift at $K$,
but are not in fact found. This sample gives $\gamma = 2.0 \pm 0.7$, quite
consistent with
the values from the CFRS.

It should be noted that a very similar calculation was carried out by Lilly et
al (1995b).
They found that  $\langle V/V_{max} \rangle$
for colour-selected ellipticals was not significantly below 0.5, but for a {\em
non-evolving model}.
As mentioned previously, however,
passive evolution has now been observationally established for cluster
ellipticals,
so a non-evolving model does not represent a viable physical picture.

\begin{figure}
\centerline{
\psfig{figure=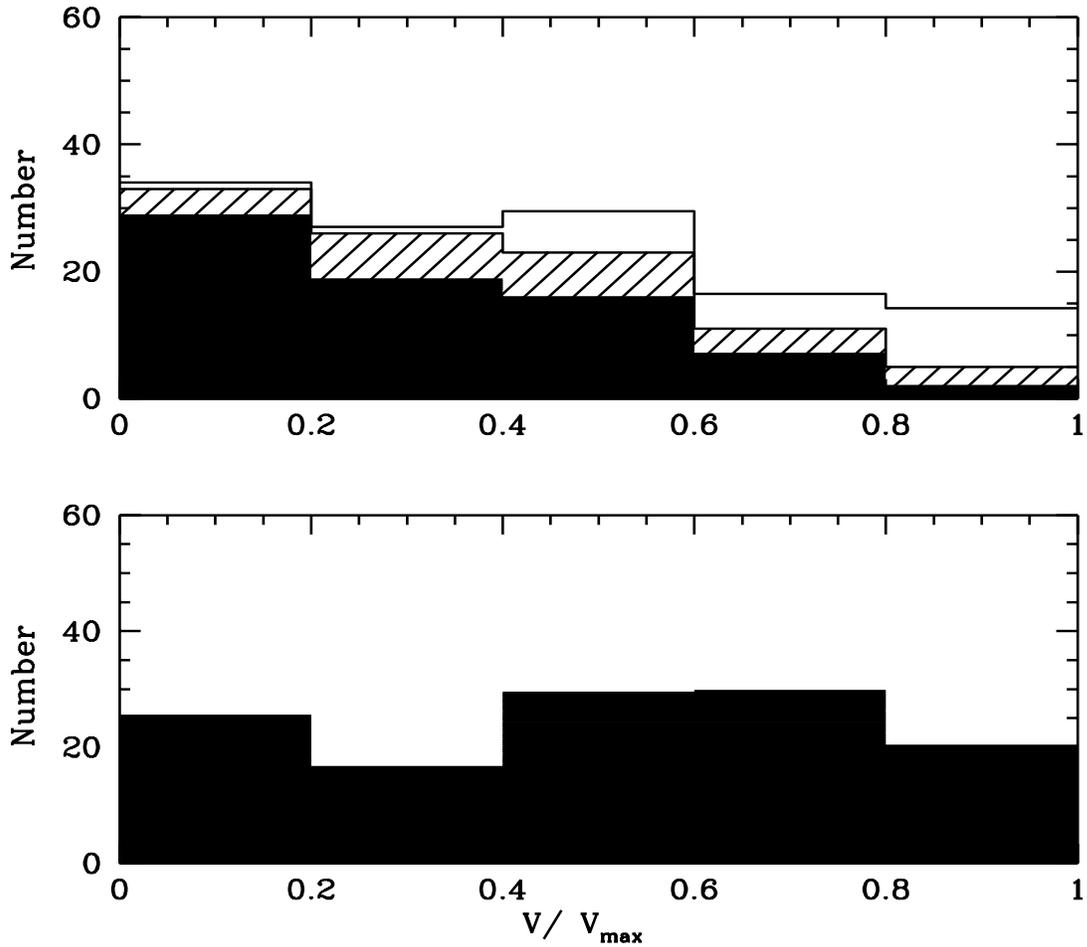,width=15cm,height=15cm}
}
\caption{\label{fig3}
{\em upper panel:} The $V/V_{max}$ distribution for
early-type galaxies selected from the CFRS catalogue. Explanations of the
histogram shadings
are as in figure 2. {\em lower panel:} The resulting $V/V_{max}$ distribution
when density
evolution of the form $F=(1+z)^{-1.5}$ is applied to the full (incompleteness
corrected) sample of 150 early-type
galaxies.}
\end {figure}

\section {Sensitivity of the Results to Stellar Models, Sample Selection,
Incompleteness and Photometric Error}

In this section, we describe a series of tests carried out to explore the
sensitivity of
our results to the various modelling and sample selection procedures that we
have adopted.

\begin {itemize}

\item {\em Stellar Population Models.}
The standard model of passive evolution adopted in this paper is a
single star formation burst of duration 0.1~Gyr at redshift $z_f=5$ with a
Salpeter
IMF ($x=1.35$ in equation 1). Higher formation redshifts or a longer duration
(up to 1~Gyr) of the initial starburst lead to nearly identical predictions.
If  a more recent formation epoch is assumed, the passive
evolution model is even more strongly ruled out since galaxies then
fade more rapidly between $z=1$ and $z=0$ (this more than compensates for the
fact that
high $z$ galaxies are also somewhat bluer in such a model). The results are
also only weakly sensitive to
changes in the IMF slope. Increasing $x$ leads to slower luminosity evolution
because low-mass stars evolve less rapidly than high-mass stars (see equation
1). However, even
for an extreme slope of $x=2.5$, we find that $\langle V/V_{max} \rangle$ only
increases
by 1.6\% to 0.417.
Finally, we note that
using alternative population synthesis models by Worthey (1994) or Tantalo et
al. (1996) would lead to results similar to those presented here.

\item {\em The Colour Threshold.} The 50\% solar model is our canonical
boundary for separating
early from late-type galaxies. For comparison, Table 1 shows how our results
are affected if
redder or bluer boundaries are adopted. Although changing the colour threshold
alters
the relative fractions of elliptical/S0s and spirals included in the sample
quite
substantially, the derived values of $\langle V/V_{max} \rangle$ vary very
little.

\item {\em Field-to-Field Variations.} Large-scale structures such as voids
or great walls can cause significant fluctuations in the number of galaxies
in certain redshift bins. This might cause us to underestimate the
sampling variance of $\langle V/V_{max} \rangle$. To test for this effect, we
analyze
separately each of the 5 fields that make up the
CFRS survey. These  fields are well enough separated so that structure should
not
correlate between them.
The field-to-field fluctuations in $\langle V/V_{max}\rangle$ scatter about the
mean
with  $\chi^2=6.54$ for four degrees of freedom, showing that any effect is
small.

\item {\em Incompleteness.} So far we have quoted results for
samples corrected for incompleteness as described in section 2.
We have also experimented with an alternative procedure that takes into account
the possibility
that the surveys may be systematically biased against including early-type
galaxies at redshifts
close to 1, simply because at $z > 0.7-0.8$ it becomes substantially more
difficult to
determine redshifts in the absence of strong emission lines.
Since $V$ and $I$-band photometry are available for all galaxies, even those
without redshifts, one
can evaluate the {\em maximum redshift} $z_{max}$ for which each
``unidentified'' galaxy would still lie
above our colour threshold and be classified as early-type. If $z_{max} > 0.7$,
we
assign $z_{max}$ to be the redshift of the unidentified galaxy.
If $z_{max} < 0.7$, we assign a redshift as before. In practice, what this
means is that all unidentified galaxies that {\em could} be early-type galaxies
at redshifts greater
than 0.7 are assigned their maximum possible value of $V/V_{max}$. Even this
extreme procedure only raises
$\langle V/V_{max} \rangle$ to  $0.451$ for the full CFRS sample, implying
$\gamma=1.0 \pm 0.3$ and
a factor 2 decrease in the number density of early-type galaxies by redshift 1,
rather than a factor 3.

\item {\em $V/V_{max}$ for early-type galaxies limited at I=21.5.}
As a further check, we repeat our analysis for
the subsample of early-type galaxies limited at $I=21.5$. There are then 76
galaxies
with redshifts and the
incompleteness correction adds only 4 extra objects. This sample gives
$\gamma=1.2 \pm 0.6$, in good agreement with the results for the fainter
sample.

\item {\em The Effect of Photometric Errors.} We model
the effect of photometric errors on our
results by introducing a random error in the $V$ and $I$ magnitudes that
increases for
galaxies with fainter apparent magnitudes. If the maximum errors are less than
0.4 mag, there is no significant change in the results. If the photometric
errors
are bigger than this,  $\langle V/V_{max} \rangle$ rises above its true value.
This is because of the number of blue galaxies increases at fainter
apparent magnitudes, so photometric errors have the net effect of scattering
more
faint galaxies into the early-type colour window than out of it.
The low values of $\langle V/V_{max} \rangle$ which we obtain for the CFRS and
Hawaii
data are thus  {\em overestimates} of the true values if photometric errors are
significant.

\end {itemize}

\section {Conclusions}
We conclude that uncertainties in the stellar population models, sample
selection
and incompleteness corrections are insufficient to reconcile the observations
with
the standard picture. Early-type galaxies selected from the Canada France
Redshift
Survey according to $V-I$ colour have a luminosity and redshift distribution
that
is inconsistent with the hypothesis that they have all been
evolving passively since a redshift of 1. It appears that by $z=1$, two thirds
of nearby early-type galaxies have dropped out of the sample, either because
they
are star-forming and no longer have colours consistent with an old stellar
population,
or because they have broken up into several pieces too faint to be included in
the sample.
This evolution may apply exclusively to the elliptical galaxy population,
exclusively
to the S0 galaxy population, or to both populations equally. We cannot use
colour alone
to separate ellipticals from S0s.

In recent hierarchical models of galaxy formation
(Kauffmann, White \& Guiderdoni 1993, Baugh et al 1996),
galaxy disks form as gas cools and condenses at
the centres of dark matter halos, while
elliptical galaxies form when two disk galaxies
merge.
If no further gas cools onto the elliptical, its stellar population will fade
and
it will evolve passively until the present day. In such models, ellipticals
form by mergers {\em at all redshifts}. The {\em typical} formation redshift of
an elliptical is higher in a rich cluster than in the field, simply because
galaxy-sized
perturbations collapse earlier in dense environments. The typical formation
redshift
of ellipticals also depends sensitively on the cosmological parameters,
in particular on the density parameter $\Omega$ and the normalization
parameter $\sigma_8$ (sometimes referred to as $b \equiv 1/\sigma_8$). Studies
of the
evolution of elliptical galaxies can thus provide important constraints on
these models.

In recent work, Kauffmann (1996) has shown that a ``standard'' CDM model with
$\Omega=1$
and $\sigma_8=0.7$ can explain the observed spread in the colours of elliptical
galaxies in clusters at redshifts between 0 and 0.6, as well as the apparent
passive
evolution of the colours of cluster ellipticals.
It is interesting that this model predicts that the global number density of
bright
ellipticals should decrease by a factor 2-3 by redshift 1, in rather good
agreement
with our results from the CFRS.
By contrast, if a lower
normalization of $\sigma_8=0.4$ is adopted, as would be appropriate for a
COBE-normalized
CDM model with a ``shape''-parameter $\Gamma=0.2$, one obtains much more rapid
evolution, with the number density of
ellipticals decreasing by a factor 25 by redshift 1.

There is also considerable spectroscopic evidence that many elliptical galaxies
have undergone recent episodes of star formation.
Charlot \& Silk (1994) attempt to quantify this using signatures
of intermediate-age stars in the spectra of ellipticals in clusters at
redshifts between
0 and 0.4. They find that the Balmer absorption line strengths increase
with redshift and they interpret this in terms of an increasing contribution of
younger stars to the
total light
(see also Barger et al 1996). It is
not possible, however, to determine from the spectra
the star formation history of the elliptical prior to
its last burst, or to assess what physical mechanism
was responsible for triggering the star formation.
In future, morphological information will be available to complement the
colours and
redshifts from the Canada-France survey. The combination of morphologies,
colours and redshifts
will help decide whether merging, star formation, or both, are
responsible for loss of early-type galaxies at high redshift. Hierarchical
formation models
suggest that the answer will be both.

\vspace{0.8cm}

\large
{\bf Acknowledgments}\\
\normalsize
We thank Olivier le F\`{e}vre and Len Cowie for making the CFRS and Hawaii data
available
to us in electronic
form and Fran\c{c}ois Hammer and the referee, Richard Ellis, for helpful advice
and discussions.
This work was carried out under the
auspices of EARA, a European Association for Research in Astronomy, and the TMR
Network on Galaxy Formation and Evolution  funded by the European Commission.

\pagebreak

\vspace {1.5cm}
\normalsize
\parindent 7mm
\parskip 8mm

{\bf Table 1:} Variation in  $\langle V/V_{max} \rangle$ for the CFRS and
Hawaii surveys
as the colour threshold is changed.
\vspace {0.3cm}

\begin {tabular} {lcccccc}
Model & \% E/S0 included & \% Sp rejected & \% E/S0 & $\langle V/V_{max}
\rangle$ & \% E/S0  & $\langle V/V_{max} \rangle $ \\
 & (Bower et al) & (Visvanathan) &  (CFRS) & (CFRS)& (Hawaii)& (Hawaii) \\
0.3 $Z_{\odot}$ & $98$ & $65$ & $21$ & $0.413 \pm .023$ & $34$ & $0.343 \pm
.045$ \\
0.4 $Z_{\odot}$ & $97$ & $78$ & $19$ & $0.408 \pm .025$ & $26$ & $0.356 \pm
.052$ \\
0.5 $Z_{\odot}$ & $88$ & $84$ & $17$ & $0.410 \pm .026$ & $19$ & $0.317 \pm
.060$ \\
0.6 $Z_{\odot}$ & $79$ & $92$ & $14$ & $0.409 \pm .028$ & $13$ & $0.326 \pm
.075$ \\
0.7 $Z_{\odot}$ & $65$ & $94$ & $12$ & $0.434 \pm .031$ & $11$ & $0.274 \pm
.080$ \\
\end {tabular}

\pagebreak
\Large
\begin {center} {\bf References} \\
\end {center}
\vspace {1.5cm}
\normalsize
\parindent -7mm
\parskip 3mm

Abraham, R.G., Tanvir, N.R., Santiago, B.X., Ellis, R.S., Glazebrook, K. \& Van
Den
Bergh, S., 1996, MNRAS, 279, L47

Arag\'{o}n-Salamanca, A., Ellis, R.S., Couch, W.J. \& Carter, D., 1993, MNRAS,
262, 764

Barger, A.J., Arag\'{o}n-Salamanca, A., Ellis, R.S., Couch, W.J., Smail, I. \&
Sharples, R.M., 1996, MNRAS, 279, 1

Baugh, C.M., Cole, S. \& Frenk, C.S., 1996, MNRAS, in press

Bender, R., Ziegler, B. \& Bruzual, G., 1996, ApJ, 463, L51

Bower, R.G., Lucey, J.R. \& Ellis, R.S., 1992, MNRAS, 254, 589

Bruzual, G. \& Charlot, S., 1996, in preparation

Charlot, S. \& Silk, J., 1994, ApJ, 432, 453

Charlot, S., 1996, in Leitherer, C., Fritze-Von Avendsleben, U. \& Hucrha, J.,
eds,
{}From Stars to galaxies, ASP Conf Series Vol. 98, in press

Cowie, L.L., Songaila, A., Hu, E.M. \& Cohen, J.G., 1996, AJ, in press

Crampton, D., Le F\`{e}vre, O., Lilly, S.J. \& Hammer, F., 1995, ApJ, 455, 96

Dickinson, M. 1995, in Buzzoni, A., Renzini, A. \& Serrano, A., eds., Fresh
Views on Elliptical
Galaxies, ASP conference series

Driver, S.P., Windhorst, R.A., Ostrander, E.J., Keel, W.C., Griffiths, R.E. \&
Ratnatunga, K.U., 1995, ApJ, 449, L23

Ellis. R.S., Smail, I., Dressler, A., Couch, W.J., Oemler, A., Butcher, H. \&
Sharples, R., 1996, ApJ, submitted

Frogel, J.A., Persson, S.E., Aaronson, M. \& Matthews, K., 1978, ApJ, 220, 75

Hammer, F., Crampton, D., Le F\`{e}vre, O. \& Lilly, S.J., 1995, ApJ, 455, 88

Kauffmann, G., 1996, MNRAS, 281, 487

Kauffmann, G., White, S.D.M. \& Guiderdoni, B., 1993, MNRAS, 264, 201

Le F\`{e}vre, O., Crampton, D., Lilly, S.J., Hammer, F. \& Tresse, L., 1995,
ApJ, 455, 60

Lilly, S.J., Le F\`{e}vre, O., Crampton, D., Hammer, F. \& Tresse, L., 1995a,
ApJ, 455, 50

Lilly, S.J., Tresse, L., Hammer, F., Crampton, D. \& Le F\`{e}vre, O., 1995b,
ApJ, 455, 108

Loveday, J., Peterson, B.A., Efstathiou, G. \& Maddox, S.J., 1992, ApJ, 390,
338

Marzke, R.O., Geller, M.J., Huchra, J.P. \& Corwin, H.G., 1994, AJ, 108, 337

Pahre, M.A., Djorgovski, S.G. \& De Carvalho, R.R., 1996, ApJ, 456, L74

Peletier, 1989, PhD thesis, Univ. Groningen

Phillips, S., 1993, MNRAS, 263, 86

Schade, D., Carlberg, R.G., Yee, H.K.C., Lopez-Cruz, O. \& Ellingson, E., 1996,
ApJ, 464, L63

Schmidt, M., 1968, ApJ, 151, 393

Tantalo, R., Chiosi, C., Bressan, A. \& Fagotto, F., 1996, A\&A, 311, 361

Tinsley, B.M. \& Gunn, J., 1976, ApJ, 302, 52

Van Dokkum, P.G. \& Franx, M., 1996, MNRAS, 281, 985

Visvanathan, N., 1992, AJ, 103, 1501

Worthey, G., 1994, ApJS, 94, 687

\pagebreak

\Large
\begin {center} {\bf Figure Captions} \\
\end {center}
\vspace {1.5cm}
\normalsize
\parindent 7mm
\parskip 8mm

{\bf Figure 1:} The division of the CFRS galaxies into early and late-types
according
to the colour-redshift criterion discussed in the text.
Solid circles represent  the ellipticals and S0s and stars represent the
spirals and irregulars.

{\bf Figure 2:} Redshift and absolute magnitude histograms for early-type
galaxies
selected from the CFRS catalogue. The solid area is for galaxies with no
detected
O[II] emission and the hatched area for galaxies with O[II] emission.
The unfilled area represents galaxies included by the incompleteness correction
described in section 2.

{\bf Figure 3:} {\em upper panel:} The $V/V_{max}$ distribution for
early-type galaxies selected from the CFRS catalogue. Explanations of the
histogram shadings
are as in figure 2. {\em lower panel:} The resulting $V/V_{max}$ distribution
when density
evolution of the form $F=(1+z)^{-1.5}$ is applied to the full (incompleteness
corrected) sample of 150 early-type
galaxies.

\end {document}